\documentclass{emulateapj}
\usepackage{graphicx,amsmath}
\usepackage[colorinlistoftodos]{todonotes}

\newcommand{\degrees}{\ensuremath{^\circ}}
\newcommand{\hst}{\textit{HST}}
\newcommand{\psr}{PSR~J1640+2224}
\newcommand{\til}{\raisebox{-0.7ex}{\ensuremath{\tilde{\;}}}}

\newcommand{\Mc}{\ensuremath{M_c}}
\newcommand{\Msol}{\ensuremath{M_\odot}}

\newcommand{\Teff}{\ensuremath{T_{\rm eff}}}
\newcommand{\Porb}{\ensuremath{P_b}}
\newcommand{\hstphot}{\texttt{HSTphot}}
\newcommand{\psrpi}{\ensuremath{\mathrm{PSR}\Large\pi}}
\newcommand{\mspsrpi}{\ensuremath{\mathrm{MSPSR}\Large\pi}}

\begin{document}

\title{Reconciling optical and radio observations of the binary millisecond pulsar \psr}

\author{
Sarah J.~Vigeland\altaffilmark{1},
Adam T.~Deller\altaffilmark{2}, 
David L.~Kaplan\altaffilmark{1},
Alina G.~Istrate\altaffilmark{1},
Benjamin W.~Stappers\altaffilmark{3},
Thomas M.~Tauris\altaffilmark{4,5}
}
\altaffiltext{1}{University of Wisconsin-Milwaukee, P.O. Box 413, Milwaukee, WI 53201 USA}
\altaffiltext{2}{Centre for Astrophysics and Supercomputing, Swinburne University of Technology, PO Box 218, Hawthorn, VIC 3122, Australia}
\altaffiltext{3}{Jodrell Bank Centre for Astrophysics, School of Physics and Astronomy, The University of Manchester, Manchester M13 9PL, UK}
\altaffiltext{4}{Max-Planck-Institut f\"ur Radioastronomie, Auf dem H\"ugel 69, D-53121 Bonn, Germany}
\altaffiltext{5}{Argelander-Institut f\"ur Astronomie, Universit\"at Bonn, Auf dem H\"ugel 71, 53121 Bonn, Germany}

\begin{abstract}
Previous optical and radio observations of the binary millisecond pulsar \psr\ 
have come to inconsistent conclusions about the identity of its companion, 
with some observations suggesting the companion is a 
low-mass helium-core (He-core) white dwarf (WD), 
while others indicate it is most likely a high-mass carbon-oxygen (CO) 
WD. 
Binary evolution models predict \psr\ most likely formed 
in a low-mass X-ray binary (LMXB) 
based on the pulsar's short spin period and 
long-period, low-eccentricity orbit, 
in which case its companion should be a He-core WD 
with mass about $0.35 - 0.39 \, \Msol$, 
depending on metallicity. 
If it is instead a CO WD, 
that would suggest the system has an unusual formation history. 
In this paper we present the first astrometric parallax 
measurement for this system from observations 
made with the Very Long Baseline Array (VLBA), 
from which we determine the distance to be $1520^{+170}_{-150}\,\mathrm{pc}$. 
We use this distance and a  
reanalysis of archival optical observations 
originally taken in 1995 with 
the Wide Field Planetary Camera 2 (WFPC2) on the \textit{Hubble Space Telescope (HST) } 
in order to measure the WD's mass. 
We also incorporate improvements in calibration, extinction model, and WD cooling models. 
We find that the existing observations are not sufficient 
to tightly constrain the companion mass, 
but we conclude the WD mass is $>0.4\,\Msol$ 
with $>90\%$ confidence. 
The limiting factor in our analysis is the low signal-to-noise ratio of the original \hst\ observations.
\end{abstract}

\section{Introduction}

Millisecond pulsars (MSPs) are valuable laboratories for studying a wide range 
of physics and astrophysics topics, including 
binary evolution \citep{champion+2008}, 
neutron star formation \citep{demorest+2010,of2016,tkf+17}, 
the equation of state of nuclear matter \citep{lattimer11}, 
and gravitation \citep{antoniadis+2013}. 
MSPs can also be used to observe low-frequency gravitational waves (GWs) 
as part of pulsar timing arrays \citep{hd83,EPTA2015,PPTA2015,NANOGrav2016}. 
These experiments rely on pulsar timing, 
whereby the times of arrival (TOAs) of pulses are fit 
to a complex timing model that incorporates the 
astrometric, spin, and binary properties of the pulsar \citep{ehm2006}. 
For some MSPs, other radio and optical observations can 
provide independent measurements of some of these parameters, which 
can be used to verify and improve pulsar timing solutions.

\psr\ is a fully-recycled MSP 
with spin period $P=3.16 \, \mathrm{ms}$. 
It is in a wide, nearly circular binary 
(orbital period $\Porb = 175 \, \mathrm{days}$, 
orbital eccentricity $e = 7.9725\times10^{-4}$) 
with a white dwarf (WD) companion \citep{lfc1996}. 
Pulsar timing observations have been able to detect 
the Shapiro delay---a relativistic time delay 
caused by the pulses' propagation 
through the companion's gravitational potential---
from which the companion's mass can be measured \citep{shapiro1964}. 
However, drawing conclusions from those data has been complicated 
by the fact that the orbital period is very close to half a year, 
resulting in degeneracies between 
some of the binary and astrometric parameters. 
Initial work by \citet{lohmer+2005} reported a low companion mass of 
$\Mc = 0.15^{+0.08}_{-0.05} \, \Msol$. 
Subsequent timing campaigns have not been able to reproduce consistent results. 
\citet{demorest+2013} reported a companion mass of $\Mc = 0.25 \pm 0.04 \, \Msol$ 
based on five years of observations taken as part of the North American Nanohertz Observatory 
for Gravitational Waves (NANOGrav). 
However, an analysis of the same data by \citet{vv2014} 
using Bayesian inference to 
perform the timing fit found a 95\% probability that the 
companion mass is above $0.3 \, \Msol$. 
Most recently, \citet{fonseca+2016} presented an even higher value of 
$\Mc = 0.6^{+0.4}_{-0.2} \, \Msol$ based on nine years of observations.

In addition, while optical observations of WDs can determine their masses and effective temperatures, 
in this case the constraints are very weak. 
Based on observations with the Palomar 200-in.\ telescope,
\citet{lcfwc1996} concluded  $\Teff = 3700 \pm 300 \, \mathrm{K}$
and $\Mc < 0.47 \, \Msol$. 
An analysis of observations made with the 
Wide Field Planetary Camera 2 (WFPC2) on the \textit{Hubble Space Telescope} (\hst) 
by \citet{lfc1996} found 
$\Teff = 4200 \pm 300 \, \mathrm{K}$ 
and $\Mc = 0.25 \pm 0.10 \, \Msol$, but 
a later analysis of the same data by \citet{hp1998} 
yielded $\Teff = 4460 \pm 1125 \, \mathrm{K}$ and 
an allowed mass range of $0.25 - 0.45 \, \Msol$. 
Both of these analyses were limited by the large uncertainty in the pulsar distance, 
which was estimated from the observed 
pulse-dispersion properties 
measure and the \citet{tc1993} model of the Galactic electron density distribution.

\begin{figure}[ht]
	\begin{center}
	\includegraphics[width=\columnwidth]{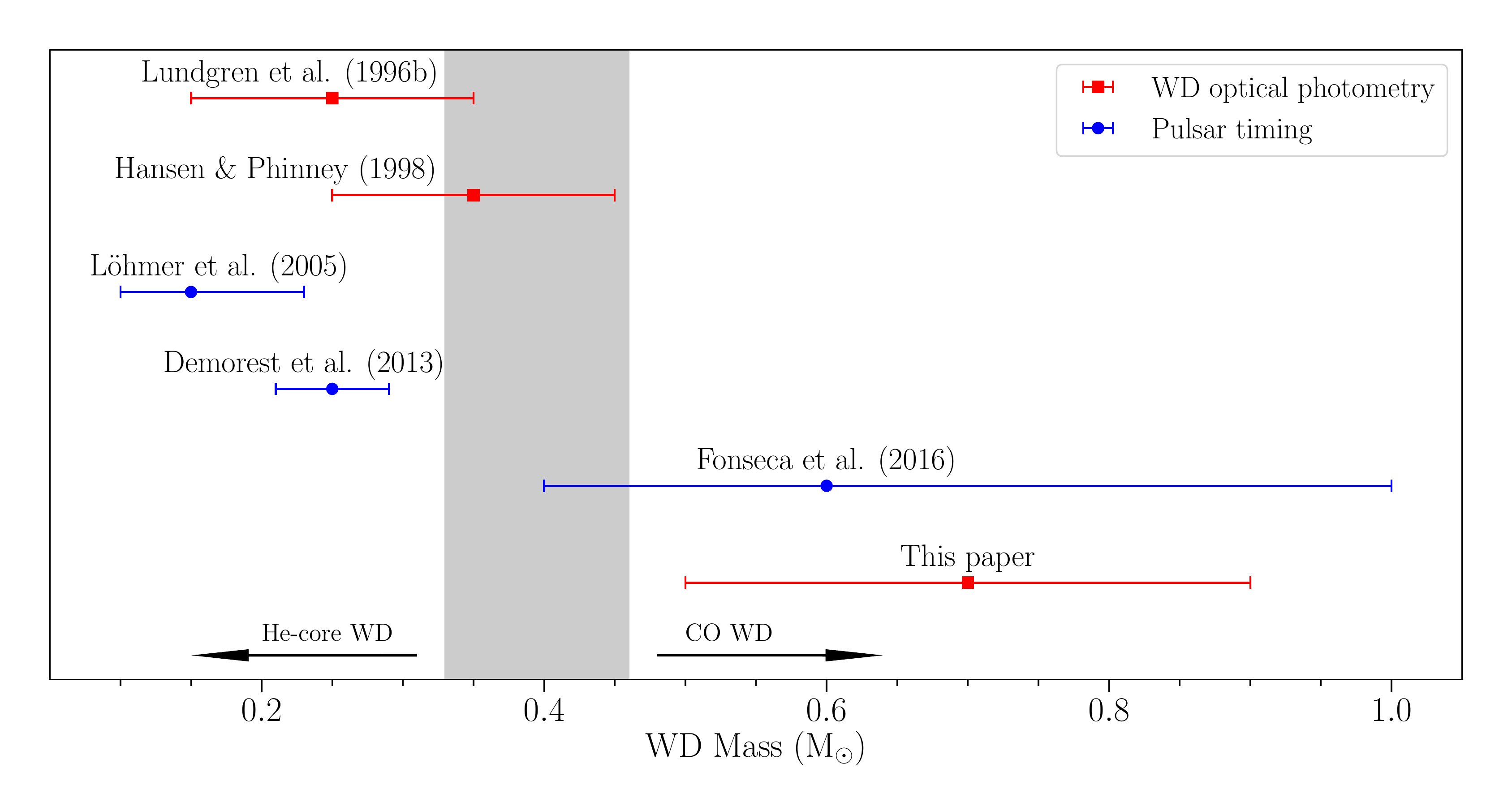}
	\caption{Published masses for the WD companion 
    of \psr. 
    We show the median values and 
    68\%-confidence intervals.
    The red squares indicate values obtained from 
    optical photometry of the WD, 
    and the blue circles indicate values 
    obtained from pulsar timing. 
    The shaded region between $0.33\,\Msol$ and $0.46\,\Msol$ 
    shows the mass cutoff 
    between He-core and CO WDs.}
	\label{fig:J1640_WDmasses}
	\end{center}
\end{figure}

\begin{figure}[ht]
	\begin{center}
	\includegraphics[width=\columnwidth]{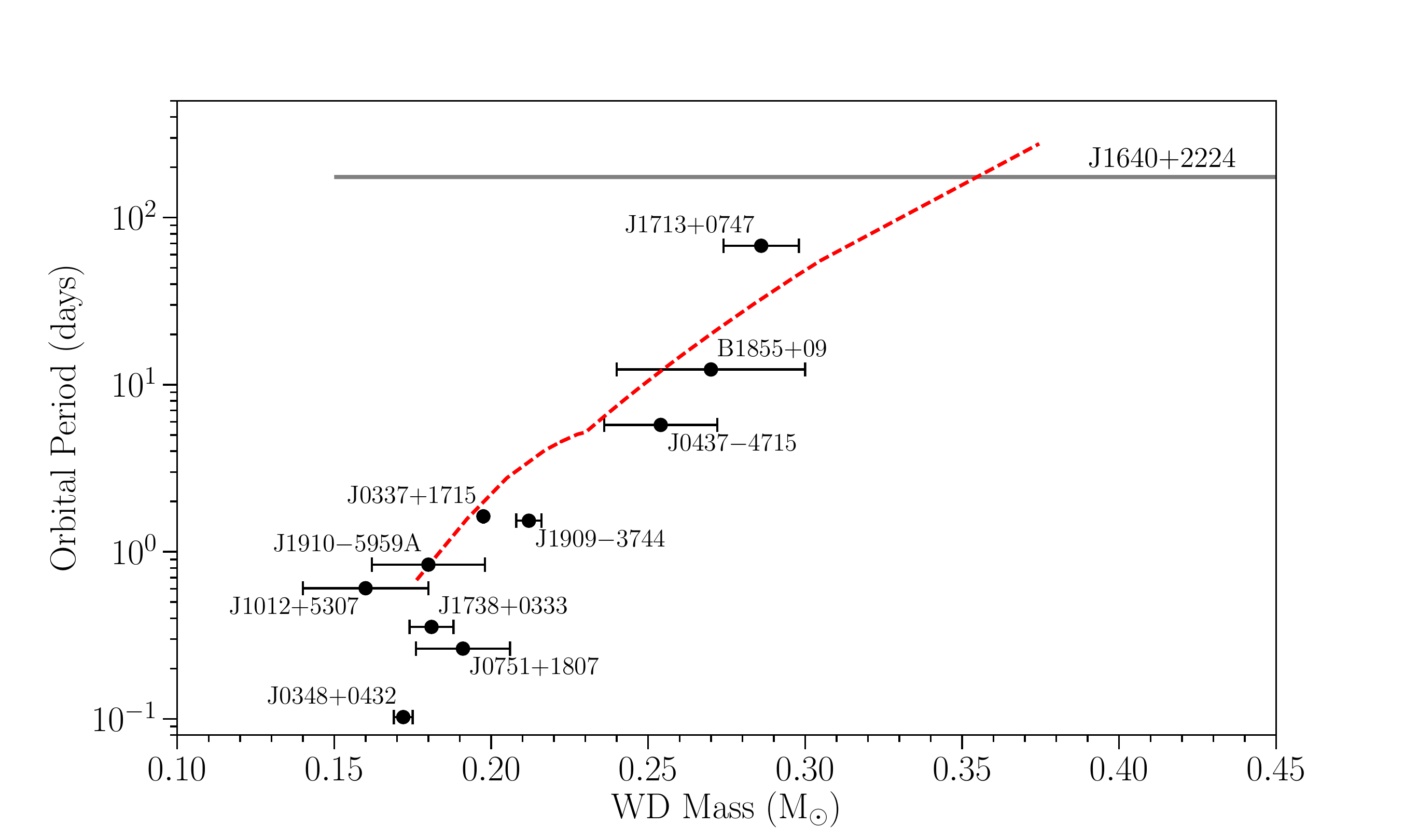}
	\caption{Orbital period and WD mass for binary MSPs with He-core WD companions. 
			The red dashed line shows the theoretical $P_\mathrm{orb} - \Mc$ relationship, 
			assuming a metallicity of $Z=0.02$ \citep{imt+16}. 
			The circles represent the measured orbital periods and WD masses, 
			with their uncertainties, for about a dozen binary MSPs. 
			All of the measured WD masses are in good agreement with the theoretical relationship.
			The gray horizontal line represents the published median values 
			for the WD companion mass of \psr, 
			which range from $0.15 \, \Msol$ \citep{lohmer+2005} 
			to $0.6 \, \Msol$ \citep{fonseca+2016}. 
			Figure modified from \citet{2017arXiv170809386S}.}
	\label{fig:PorbMwd}
	\end{center}
\end{figure}

The large range in the reported companion masses 
(see Fig.~\ref{fig:J1640_WDmasses})
has important implications 
for binary evolution models. 
The small pulsar spin period and wide, nearly circular orbit point to this system 
forming from a wide-orbit low-mass X-ray binary (LMXB; \citealt{tauris2011}). 
Models of LMXBs predict a correlation between the orbital
period and WD mass, which for this system implies the companion mass is 
$0.35 - 0.39\, \Msol$ depending on the metallicity 
(\citealt{rappaport1995, ts1999, vbjj2005, imt+16}; see Fig.~\ref{fig:PorbMwd}). 
If the companion is significantly more massive, 
as suggested by some results from pulsar timing, 
then how the system formed is uncertain. 
Most pulsars with massive WD companions formed in 
intermediate-mass X-ray binaries (IMXBs; \citealt{tauris2011}). 
However, \psr\ has a smaller spin period, longer orbital period, 
and smaller orbital eccentricity than most binary pulsars formed this way.
In particular, the combination of these characteristics makes it difficult to understand the formation of PSR~J1640+2224 if it has a CO WD companion and thus formed from an IMXB.

In this paper, we present the first astrometric parallax measurement for \psr\ 
based on observations taken with the Very Long Baseline Array (VLBA). 
We use this distance to reanalyze the 1995 \hst\ observations 
of \psr. We also improve upon the original analysis by taking advantage 
of improved photometry software, 
updated calibration of WFPC2 \citep{dolphin2000, dolphin2009}, 
a new 3D map of Galactic extinction \citep{green+2015}, 
and updated WD cooling models \citep{tbg11, bwd+11}. 
Our analysis of two other pulsars observed in the same manner, 
PSR J1022+1001 and PSR J2145$-$0750, were published in \citet{deller+2016}. 

This paper is organized as follows. 
In Sec.~\ref{sec:VLBI} we describe the very long baseline interferometry (VLBI) observations used 
to measure the astrometric parallax and transverse velocity of the system. 
In Sec.~\ref{sec:observations} we describe the optical observations 
and data reduction techniques used. 
In Sec.~\ref{sec:coolingcurves} we discuss the procedure 
used to fit the photometry data 
to the WD cooling curves to obtain the temperature and mass of the WD. 
We summarize our findings in Sec.~\ref{sec:conclusions}.

\section{VLBI Observations}
\label{sec:VLBI}

We observed \psr\ a total of 11 times with the VLBA in the period 2015 August to 2017 August under the project codes BD179 and BD192, as part of the \mspsrpi\ project\footnote{\url{https://safe.nrao.edu/vlba/psrpi/status.html}}.  The observing and data reduction strategy are similar to that used previously in \citet{deller+2016} for PSRs J1022+1001 and J2145--0750 and described in detail in the \psrpi\ catalog paper (Deller et al., in prep).  \psr\ is one of the first sources in the \mspsrpi\ program to complete observations; results for a total of 18 millisecond pulsars (including a comparison of VLBI-derived to timing-derived positions and proper motions to test frame alignment) will be available over the coming 12 months.

The VLBA's maximum recording rate of 2 gigabits per second was employed to sample 256 MHz of bandwidth in dual polarization at each antenna, spanning the range 1392.0--1744.0 MHz while avoiding regions contaminated by radio frequency interference.  The source J1594+1029 was observed once per observation to calibrate the instrumental delays and bandpass, while J1641+2257 was observed every 
5 minutes to calibrate atmospheric/ionospheric delays and phases. Around 40 minutes on the target source \psr\ was obtained per observation.

To refine the atmospheric/ionospheric phase calibration in the pulsar direction and produce the highest quality astrometry, the use of a secondary phase calibrator (preferably in the same field of view as the target pulsar) is highly desirable. A total of four suitably compact and bright background sources were identified in the same field as \psr\ during a short exploratory observation prior to the commencement of astrometry in 2015, in which the multi-field capability of the DiFX correlator used at the VLBA \citep{deller11a} was employed to inspect every known radio source with 30\arcmin\ of \psr.  The source J164018.9+221203 was the brightest source in the field (flux density $\sim$100 mJy, angular separation 12\arcmin\ from \psr) and was used for secondary phase calibration, while the other sources are used for consistency checks.
The positions used for all relevant calibrator sources are shown in Table~\ref{tab:calsources}; the in-beam source reference positions are derived from relative astrometry to J1641+2257 (reference epoch MJD 57500), using the observations presented here.

\begin{deluxetable}{l l l c}
  \tablewidth{0.95\columnwidth}
\tablecaption{VLBI calibrator sources\label{tab:calsources}}
\tablehead{
  \colhead{Source} & \colhead{Right Ascension} & \colhead{Declination} & \colhead{Reference} 
  }
\startdata
J1641+2257			& 16:41:25.227564	& 22:57:04.03284	& RFC2015a\footnote{\url{http://astrogeo.org/rfc/}} \\
J163919.7+221120	& 16:39:19.77745	& 22:11:20.0485	& This work \\
J164018.7+222917	& 16:40:18.71436	& 22:29:17.4158	& This work \\
J164018.9+221203	& 16:40:18.90807	& 22:12:03.9182	& This work \\
J164044.3+221016	& 16:40:44.38302	& 22:10:16.4713 & This work
\enddata
\end{deluxetable}

For each astrometric observation, a total of six correlated datasets were produced: one for each of the background sources, and two at the position of \psr, where one correlation employed pulsar gating \citep{deller07a} to boost sensitivity on the pulsar, and one was ungated.  The pulsar gating made use of an ephemeris derived from observations with the Lovell telescope and was used for all astrometry, while the ungated dataset was used to check the correctness of the pulsar ephemeris.  Calibration was derived and applied in the manner described in \citet{deller+2016}, and used a solution interval of 6 seconds for secondary phase calibration (frequency independent) and 90 seconds for secondary phase calibration (frequency dependent).  A position time series for each of the five in-beam sources (4 background and the pulsar) was obtained via imaging and position fitting as described in \citet{deller+2016}.  Finally, we add an additional component in quadrature to the position uncertainties at each epoch based on the angular separation between the source and the secondary phase calibrator J164018.9+221203, to account for systematic position errors resulting from the differential ionosphere.  The component added in quadrature ranged from 0.05 -- 0.2 mas depending on the source, and was calculated using empirical results from the \psrpi\ VLBI astrometric campaign containing 60 pulsars \citep[Deller et al. in prep]{deller+2016}.

\begin{figure}[htb]
	\begin{center}
    \begin{tabular}{c}
	\includegraphics[width=\columnwidth]{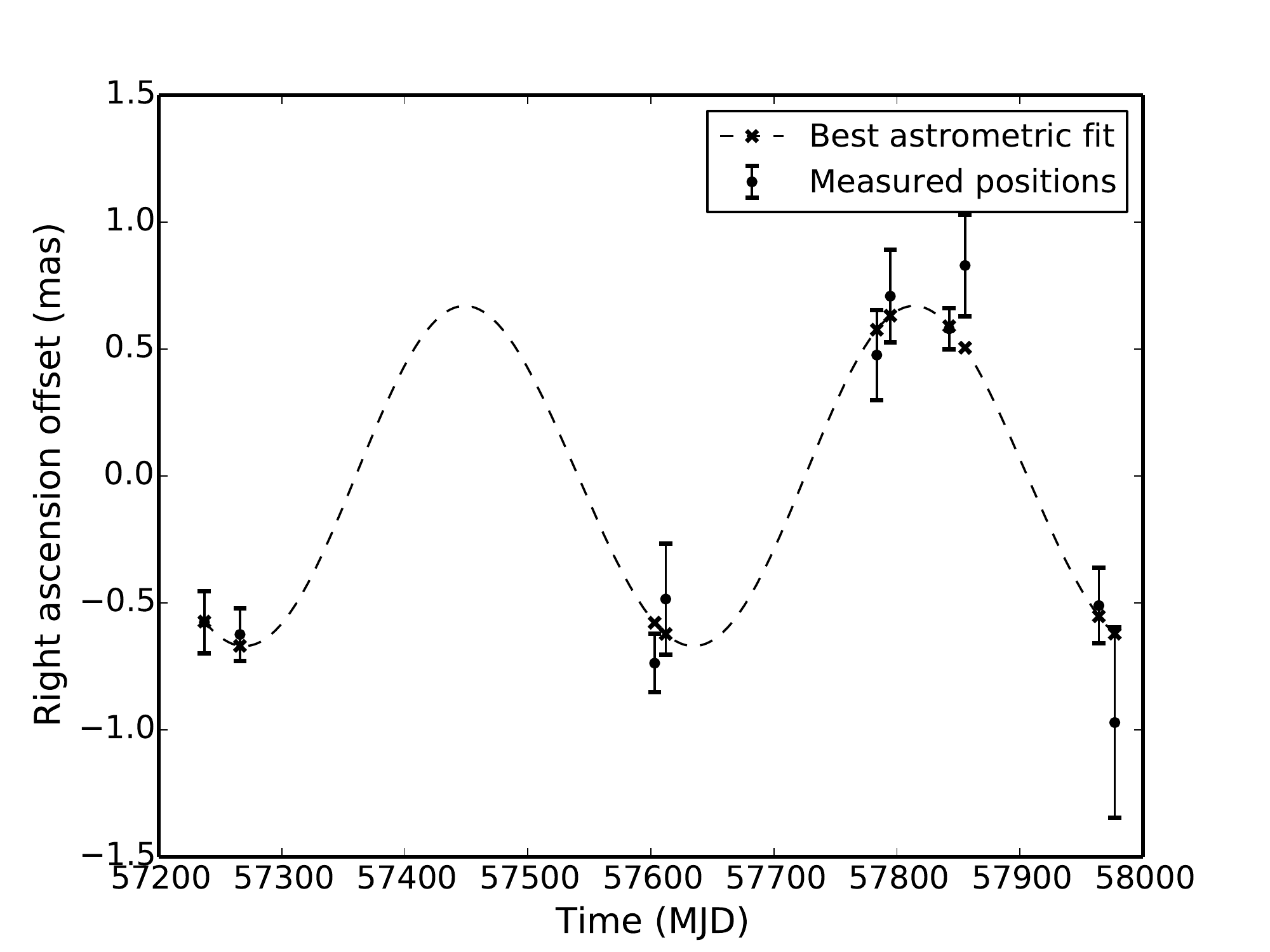} \\
    \includegraphics[width=\columnwidth]{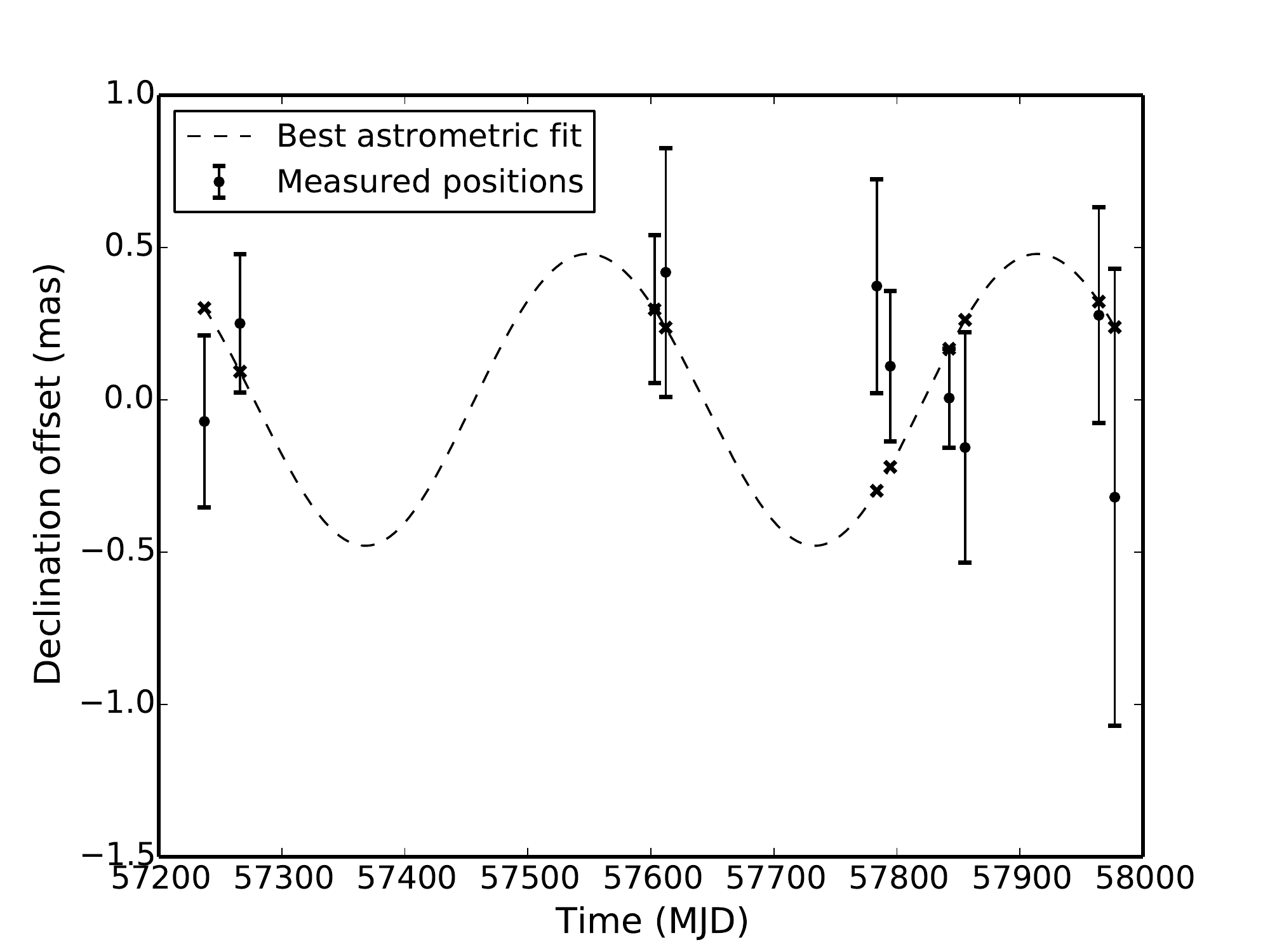}
    \end{tabular}
	\caption{Relative position offset for \psr\ in right ascension (top) and declination (bottom) 
			as a function of time, after subtracting proper motion to highlight the parallax contribution.  
			For plotting purposes, the poorly constrained orbital reflex motion has been neglected.}
	\label{fig:VLBIfit}
	\end{center}
\end{figure}

We fit each position time series for reference position, proper motion, and parallax using the linear least-squares solver \texttt{pmpar}\footnote{\url{https://github.com/walterfb/pmpar}}.  For \psr, we performed fits using the affine-invariant Markov chain Monte Carlo (MCMC) ensemble sampler \texttt{emcee} \citep{fmhlg13},  solving additionally for orbital inclination $i$ and longitude of ascending node $\Omega$ in order to account for orbital reflex motion.  We used uniform priors for all parameters, excluding only values $i\leq14$\degrees, which are ruled out by the mass function obtained by pulsar timing for a companion mass $\leq1.4$ \Msol, assuming the pulsar mass is $\geq1.2$ \Msol. We also cross-checked our results using bootstrap sampling in the manner described in \citet{deller+2016}, obtaining consistent values and uncertainties.  The results 
(with 68\% credible intervals) 
are shown in Table~\ref{tab:VLBIfit} and plotted in Figure~\ref{fig:VLBIfit}. No significant constraints could be placed on $i$ or $\Omega$ based on the orbital reflex motion, which was small compared to the typical position uncertainty in a single epoch.  The unknown reflex motion did, however, lead to larger uncertainties on the parallax and proper motion, and so if the inclination in particular could be better constrained, the distance and transverse velocity uncertainties for \psr\ could be further reduced.  The VLBI-estimated position and proper motion are consistent at the 1-$\sigma$ level with the 9-year NANOGrav results presented by \citet{mnf+16}.

As an additional cross-check, we inspected the astrometric fits for the three other background sources not used for secondary phase calibration.  In the absence of calibration errors, if all sources are distant background objects that do not exhibit structural variations, then their fitted proper motions and parallaxes should be consistent with zero.  In most cases, the fitted values are consistent with zero, and in almost all cases they are smaller than the uncertainties on the corresponding parameters for \psr.  The exceptions were J163919.7+221120, with a significant measurement of $\mu_\alpha = 0.17 \pm 0.05$ mas yr$^{-1}$, and J164018.7+222917 with $\mu_\alpha = 0.26 \pm 0.12$ mas yr$^{-1}$.  Both of these sources have a larger angular separation to the secondary phase calibrator than does \psr, meaning that any systematic errors in the fitted parameters for \psr\ should be smaller than these values.  If structural evolution in J163919.7+221120 or  J164018.7+222917 played a significant role in producing these offsets, then the maximum expected value of any systematic errors for \psr\ would be even smaller.

\begin{deluxetable}{l c}
  \tablewidth{0.9\columnwidth}
\tablecaption{VLBI Astrometric Parameter Fits\label{tab:VLBIfit}}
\tablehead{
  \colhead{Parameter} & \colhead{Value}
  }
\startdata
R.A. (J2000) & 16:40:16.74587\,$\pm\,0.00007$ \\
Decl. (J2000) & 22:24:08.764\,$\pm\,0.001$ \\
Position epoch (MJD) & 57500 \\
R.A. offset\tablenotemark{A} (mas) & $-29985.23 \pm 0.07$ \\
Decl. offset\tablenotemark{A} (mas) & $724845.93 \pm 0.12$ \\
$\mu_\alpha$ ($\mathrm{mas}\,\mathrm{yr}^{-1}$) & $2.19 \pm 0.08$ \\
$\mu_\delta$ ($\mathrm{mas}\,\mathrm{yr}^{-1}$) & $-11.28 \pm 0.14$ \\
Parallax (mas) & $0.66 \pm 0.07$ \\
Distance (pc) & $1520^{+170}_{-150}$ \\
$v_T$ ($\mathrm{km}\,\mathrm{s}^{-1}$) & $82^{+11}_{-9}$
\enddata
\tablenotetext{A}{Relative to the reference position for J164018.9+221203.}
\end{deluxetable}

\section{Optical Observations}
\label{sec:observations}

\begin{figure}[htb]
	\begin{center}
	\includegraphics[width=\columnwidth]{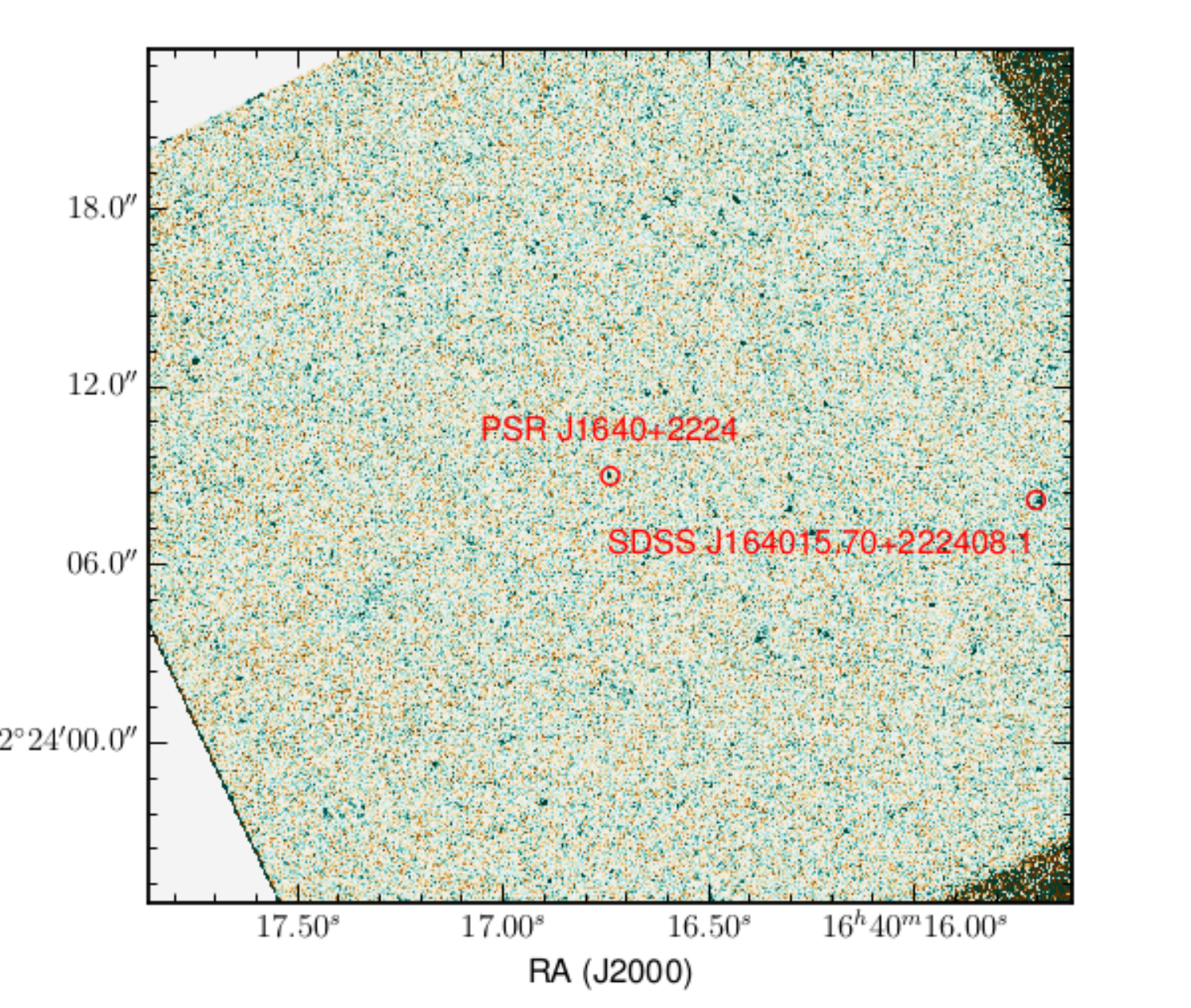}
	\caption{Drizzled \citep{fh2002} image of \psr\ from \hst\ observations by \citet{lfc1996} 
			using WFPC2. The central red circle indicates the position of the pulsar based on 
			pulsar timing observations, corrected to the observation epoch. 
			The circle's radius of 0\farcs3 reflects the absolute 
			astrometric uncertainty. The optical source, located inside the red circle, is consistent 
			with the pulsar. For reference, the other red circle 
			indicates a source from the Sloan Digital Sky Survey \citep{SDSS}.  
			The astrometric agreement allows us to conclude that we have 
			in fact identified the counterpart to \psr\ (cf.\ \citealt{lfc1996}).}
	\label{fig:J1640image}
	\end{center}
\end{figure}

\psr\ was observed with the Wide Field and Planetary Camera 2 (WFPC2) aboard 
\hst\ in October 1995 as part of a program to observe WD companions of six MSPs. 
The source was imaged with the Planetary Camera (PC) detector. 
Images were made with the F555W ($V$-band) and F814W ($I$-band) filters with 
exposure times of 1000 s and 800 s, respectively. 
Figure~\ref{fig:J1640image} shows a drizzled \citep{fh2002} combined image 
downloaded from the Hubble Legacy Archive. The absolute astrometry has been 
improved to 0\farcs3, compared to 0\farcs8 from \citet{lfc1996}.

Data reduction was done with \hstphot\ (version 1.1; \citealt{dolphin2000}). 
We masked bad pixels, estimated the sky level, and masked cosmic rays by comparing 
exposure pairs. We used point-spread function (PSF) fitting photometry using the revised 
calibration of \citet{dolphin2009}. Since the fields were relatively sparse, we did not allow 
\hstphot\ to determine the PSF residual and we used the default aperture corrections. 
There were no significant differences in the final magnitudes 
when we varied the processing steps. 
We found $25.52 \pm 0.12 \, \mathrm{mag}$ in the F555W filter, 
and $24.62 \pm 0.16 \, \mathrm{mag}$ in the F814W filter. 
Our F814W magnitude is the same as the value reported by \citet{lfc1996}, 
but our F555W magnitude is $0.5 \, \mathrm{mag}$ brighter 
(a 1.7-sigma difference).

The extinction and its uncertainty were obtained from 
the 3D Galactic dust map by \citet{green+2015}, which gives the reddening $E(B-V)$ 
along the line of sight as a function of distance, 
using the VLBI-measured distance. 
We converted from $E(B-V)$ to the extinctions in each band using 
$R_V=3.1$ and the extinction coefficients $A_\lambda$ from \citet{girardi+2008}, 
with an additional 15\% reduction in $A_\lambda$ 
based on the revised calibration in \citet{sf2011}, 
and found $A_V = 0.12 \pm 0.03$.

\section{WD Atmosphere Fitting}
\label{sec:coolingcurves}
We fit the photometry 
using both hydrogen-atmosphere (DA) and helium-atmosphere (DB) models 
from \citet{tbg11} and \citet{bwd+11}, 
respectively\footnote{Also see \url{http://www.astro.umontreal.ca/{\til}bergeron/CoolingModels/}.}.
These models tabulate synthetic photometry integrated throughout the
\hst/WFPC2 filter passbands in the Vega system (like
\texttt{hstphot}) for a range of effective temperatures, masses, and radii. 
To simplify our analysis, we only used WD atmosphere models 
for a single reference mass $M_\mathrm{ref}$. 
We compared the model magnitudes 
with the observed apparent magnitudes according to
\[
	m = M + 5 \log_{10} \left( \frac{d}{10 \, \mathrm{pc}} \right) + A_\lambda - 5 \log_{10} \left[ \frac{R_c(\Teff, M_c)}{R_\mathrm{ref}} \right] \,,
\]
where $m$ is the observed apparent magnitude, 
$M$ is the absolute magnitude from the WD models, 
$d$ is the distance, 
$A_\lambda$ is the extinction in a particular band, 
$R_c$ is the WD radius, 
\Teff\ is the effective temperature, 
and $M_c$ is the WD mass. 
$R_\mathrm{ref}$ is the radius of a WD 
with temperature \Teff\ and 
whose mass is the reference mass $M_\mathrm{ref}$. 
The final term in this expression incorporates 
the dependence on the WD mass and radius by adjusting 
the magnitudes based on the ratio between the WD radius 
for a particular effective temperature and WD mass [$R_c(\Teff, M_c)$] 
and the radius for that temperature and the model mass [$R_\mathrm{ref}$]. 
This simplification introduces an error of $<0.1 \, \mathrm{mag}$, 
which is smaller than the photometric uncertainties.

We used two sets of models to determine the WD radius. 
Since the DA and DB models assume a carbon-oxygen (CO) core, 
we only used the WD radius from these models for $\Mc \geq 0.4 \, \Msol$, 
where the assumption of a CO core is likely correct. 
For $\Mc < 0.4 \, \Msol$, we used low-mass He-core WD models from \citet{imt+16} 
to determine the radius as a function of mass and temperature.

We performed the model fits with the 
affine-invariant Markov chain Monte Carlo (MCMC) ensemble sampler \texttt{emcee} \citep{fmhlg13}. 
We used a uniform prior on \Teff\ between $4000$ and $10\,000 \, \mathrm{K}$, 
and a Gaussian prior on $A_\lambda$ with mean and standard deviation 
determined from the \citet{green+2015} 3D Galactic dust map for this distance and line of sight. 
We used a Gaussian prior on the parallax centered around the VLBI-measured value 
and a uniform prior on the WD mass between $0.2$ and $1.2 \, \Msol$.

\begin{figure}[ht]
	\begin{center}
	\includegraphics[width=\columnwidth]{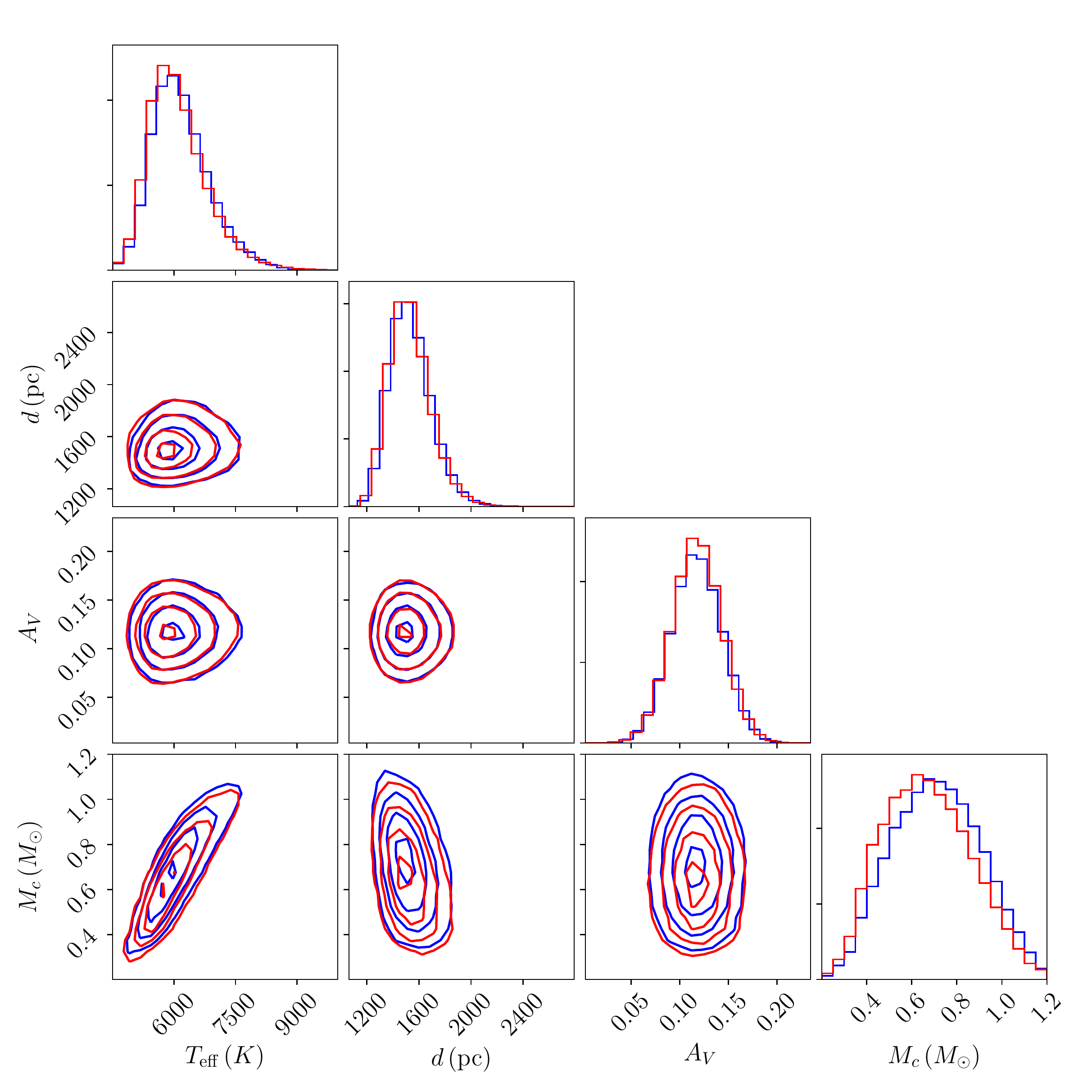}
	\caption{Joint two-dimensional posterior probability distributions 
			and marginalized one-dimensional posterior probability distributions 
			from fitting the \hst\ photometry to WD cooling models. 
			We use the VLBI-measured parallax as a prior on the distance, 
			and a Gaussian prior on the extinction $A_V$ with mean and standard deviation 
			taken from the \citet{green+2015} 3D galactic dust map. 
			We show fits to both DA model WDs (blue curves) and DB model WDs (red curves). 
			Best-fit parameter values and uncertainties are listed in Table~\ref{tab:mcmc_fit}.}
	\label{fig:J1640mcmc}
	\end{center}
\end{figure}

Table~\ref{tab:mcmc_fit} lists the best-fit values, and 
Fig.~\ref{fig:J1640mcmc} shows the posterior distributions 
obtained using DA and DB WD models. 
We find an effective temperature of 
$\Teff = 6090^{+780}_{-590} \, \mathrm{K}$ and 
$\Teff = 6000^{+790}_{-570} \, \mathrm{K}$ for DA and DB WDs, respectively. 
These values are significantly higher than the value reported by \citet{lfc1996}, 
but they are consistent with those in \citet{hp1998}. 
For the WD mass, we find 
$\Mc = 0.71^{+0.21}_{-0.20} \, \Msol$ with DA models 
and $\Mc = 0.66^{+0.21}_{-0.19} \, \Msol$ with DB models, 
and our results indicate $\Mc > 0.4 \, \Msol$ with $>90\%$-confidence 
for both DA and DB models. 
This suggests the companion is most likely a CO WD.

\begin{deluxetable}{l c c}
  \tablewidth{0.95\columnwidth}
\tablecaption{WD Parameter Fits\label{tab:mcmc_fit}}
\tablehead{
  \colhead{Parameter} & DA WD & DB WD} \\
\startdata
\Teff\ (K) & $6090^{+780}_{-590}$ & $6000^{+790}_{-570}$  \\
$d$ (pc) & $1520^{+180}_{-140}$ & $1520^{+170}_{-140}$ \\
$A_V$ & $0.12^{+0.03}_{-0.03}$ & $0.12^{+0.03}_{-0.03}$\\
$R_c$ ($10^{-2} R_\odot$) & $1.12^{+0.26}_{-0.23}$ & $1.18^{+0.26}_{-0.24}$ \\
$M_c$ ($\Msol$) & $0.71^{+0.21}_{-0.20}$ & ${0.66^{+0.21}_{-0.19}}$ \\
$\chi^2$ & $1.33$ & $1.54$
\enddata
\tablecomments{Values are the median and 68\% confidence intervals from the marginalized posterior distributions.}
\end{deluxetable}

\section{Discussion and Conclusions}
\label{sec:conclusions}

Here we present measurements of the parallax and transverse velocity 
of \psr\ from VLBI astrometry, 
including the first precise distance measurement for this system. 
We measured an annual geometric parallax of $0.66 \pm 0.07 \, \mathrm{mas}$, 
which yields a distance of $1520^{+170}_{-150} \, \mathrm{pc}$.
We used this distance measurement to reanalyze the 1995 \hst\ observations of \psr\ 
in order to infer the WD companion's mass. 
We were also able to improve upon the original analysis 
by taking advantage of improvements in \hst\ calibration, updated WD cooling models, 
and a new 3D model of Galactic dust.

The original analysis by \citet{lfc1996} 
identified the companion as a low-mass He-core WD 
with $\Teff = 4200 \pm 300 \, \mathrm{K}$ and $\Mc = 0.25 \pm 0.10 \, \Msol$. 
Our reanalysis found a significantly higher temperature of 
$\Teff = 6090^{+780}_{-590} \, \mathrm{K}$ and 
$\Teff = 6000^{+790}_{-570} \, \mathrm{K}$ for DA and DB WDs, respectively. 
We also found a higher companion mass than previously reported, 
$\Mc = 0.71^{+0.21}_{-0.20} \, \Msol$ with DA models 
and $\Mc = 0.66^{+0.21}_{-0.19} \, \Msol$ with DB models. 
Our analysis finds $\Mc > 0.4 \, \Msol$ with $>90\%$-confidence for both DA and DB models, 
indicating the companion is mostly likely a CO WD, 
rather than a He-core WD as formation models predict.

Identifying the WD companion as either a He-core WD or CO WD 
has implications for NS-WD binary formation models. 
The low spin period indicates \psr\ has been fully recycled, 
and that combined with its low orbital eccentricity implies it formed 
in a LMXB, in which case its companion should be a He-core WD. 
If that is the case, a precise measurement of the WD mass 
can be used to constrain the theoretical $\Porb-\Mc$ 
relationship for wide-orbit LMXBs. 
There are only a dozen observations known to verify this relationship since relatively few pulsar binaries have precise measurements of the companion mass, especially in wide systems with $P_b > 20\;{\rm days}$ \citep{tvh14}. 
Furthermore, there are a number of factors that may affect the WD mass, 
including the convective mixing-length parameter 
and the metallicity of the WD progenitor star
\citep{ts1999, stairs+2005, sl2012, imt+16}.

If the companion is instead a CO~WD, then the formation of this system is a puzzle. 
Most pulsars with CO~WD companions formed in IMXBs, 
which typically yield partially-recycled MSPs with relatively  
long spin periods ($P>10 \, \mathrm{ms}$) and short orbital periods 
($\Porb<40\,\mathrm{days}$). 
However, there are exceptions. 
The binary pulsar J1614$-$2230 has a massive CO WD companion and a 
spin period of $3.15 \, \mathrm{ms}$, which is comparable to \psr, 
although its orbital period is significantly shorter 
($\Porb=8 \, \mathrm{days}$; \citealt{tlk2011,tlk2012}). 
Another possibility is that 
\psr's formation involved dynamical encounters, 
as is believed to be the case for 
PSR J1903+0327 \citep{champion+2008}, 
PSR J1835$-$3259A \citep{decesar+2015}, and 
PSR J1024$-$0719 \citep{kaplan+2016, bassa+2016}. 
However, PSR J1903+0327 and PSR J1835$-$3259A both have 
highly eccentric orbits\footnote{The orbital eccentricity of PSR J1024$-$0719 is not well constrained 
due to the extremely long orbital period.} 
whereas \psr\ is in a nearly circular orbit.

Measurements of the WD mass are limited by the low S/N of the original observations, 
as well as having observations in only two filters ($V$-band and $I$-band). 
Higher precision optical observations in three or more filters 
could be used to get a more precise mass measurement. 
Additionally, long-term high-cadence timing observations of \psr\ 
may be able to break the degeneracies in the timing model 
between the astrometric and binary parameters, 
allowing for the companion mass to be measured via the Shapiro delay. 

\acknowledgements

We thank Saul Rappaport and Joris Verbiest for 
helpful comments and suggestions. 
SJV~and DLK~are funded by the NSF Physics Frontiers Center award number 1430284. 
AGI~acknowledges support from the NASA Astrophysics Theory Program
through NASA grant NNX13AH43G. 
Access to the Lovell Telescope and pulsar research at the Jodrell Bank Centre for Astrophysics 
is supported through an STFC consolidated grant. 
The Long Baseline Observatory is a facility of the National Science Foundation 
operated under cooperative agreement by Associated Universities, Inc. 
Based on observations made with the NASA/ESA \textit{Hubble Space Telescope},
and obtained from the Hubble Legacy Archive, which is a collaboration
between the Space Telescope Science Institute (STScI/NASA), the Space
Telescope European Coordinating Facility (ST-ECF/ESA) and the Canadian
Astronomy Data Centre (CADC/NRC/CSA).

{\it Facilities:} \facility{HST (WFPC2)}, \facility{VLBA}

\bibliographystyle{aasjournal}
\bibliography{master}

\end{document}